\documentclass[10pt,aps,prl,superscriptaddress,nopacs,amsmath,amstex,amssymb,citeautoscript,longbibliography,floatfix,twocolumn]{revtex4-2}
\usepackage{graphicx,amsmath,amssymb,amsfonts,color}
\usepackage{nicefrac}
\usepackage{multirow, makecell, ragged2e}
\usepackage{physics}
\usepackage{bm}
\usepackage{float,lipsum}
\usepackage{algpseudocode}
\usepackage{algorithm}
\usepackage{placeins}
\usepackage{ulem}

\usepackage[colorlinks,bookmarks=true,citecolor=blue,linkcolor=red,urlcolor=blue]{hyperref}

\usepackage[nameinlink,noabbrev]{cleveref}



\usepackage{graphicx}
\usepackage{subfigure}
\usepackage{adjustbox}
\usepackage{bm}
\usepackage{color}
\usepackage{braket}
\usepackage{standalone}
\usepackage{multirow}
\usepackage{tikz}
\usepackage{mathrsfs}
\usepackage{dsfont}
\usepackage{comment}
\usepackage{amsmath}

\usepackage{orcidlink}
\usepackage{subcaption}

\usepackage{graphicx,amsmath,amssymb,amsfonts, color}
\usepackage{nicefrac}
\usepackage{multirow, makecell, ragged2e}
\usepackage{bm}
\usepackage{float,lipsum}
\usepackage{algpseudocode}
\usepackage{algorithm}

\newcommand{\be}{\begin{equation}}
\newcommand{\ee}{\end{equation}}

\begin{document}

\title{Topological Anderson Superconductor in Landau Levels}
\author{Uddalok Nag}
\affiliation{Department of Physics, 104 Davey Lab, The Pennsylvania State University, University Park, Pennsylvania 16802, USA}
\affiliation{Center for Theory of Emergent Quantum Matter, The Pennsylvania State University, University Park, Pennsylvania 16802, USA}
\author{Koji Kudo}
\affiliation{Department of Physics, Kyushu University, Fukuoka 819-0395, Japan}
\author{J. K. Jain\orcidlink{000-0003-0082-5881}}
\affiliation{Department of Physics, 104 Davey Lab, The Pennsylvania State University, University Park, Pennsylvania 16802, USA}
\affiliation{Center for Theory of Emergent Quantum Matter, The Pennsylvania State University, University Park, Pennsylvania 16802, USA}
\affiliation{Lodha Theoretical Physics Institute, 17th Floor, Lodha NCP Supremus, Wadala (E), Mumbai 400037, India}

\begin{abstract}
There is much interest in two-dimensional superconductors with odd integer values of the Bogoliubov-de Gennes (BdG) Chern number $C_{\rm BdG}$, often termed topological superconductors, as these are predicted to harbor Majorana zero modes. Because a magnetic field produces Landau levels (LL) with nonzero Chern numbers, it is natural to ask if a conventional two-dimensional superconductor of spinful fermions with Rashba spin-orbit coupling can be driven topological by application of a magnetic field. Self-consistent solutions of the BdG equations for an on-site attractive interaction produce a phase diagram consisting almost everywhere of superconducting states with even integer values of $C_{\rm BdG}$, transitioning from one value to another as parameters such as the chemical potential are varied. We predict that introducing Anderson-type on-site disorder, which broadens as well as mixes LLs, produces, in the transition region, states wherein topological superconductivity percolates. This topological Anderson superconducting state provides another example where disorder can produce topologically nontrivial states. 
\end{abstract}
\graphicspath{{./Figures/}}

\maketitle

\section{Introduction}
In topological superconductors, the Bogoliubov-de Gennes (BdG) quasiparticle states posses a non trivial band topology that produces an odd integer value of the BdG Chern number $C_{\rm BdG}$ (defined below). A canonical example is the chiral $p_x+ip_y$ superconductor, whose nontrivial topology gives rise to chiral Majorana modes at the boundary and Majorana zero modes bound to vortices~\cite{Read00,Ivanov01,Stone2004,Kraus2009,sato2017}.
It has been proposed that such excitations obey non-Abelian braid statistics and can provide a route to fault-tolerant quantum computation~\cite{Kitaev03,Stone2006,Oshikawa2007,Nayak2008,Sau2010,Stanescu2010,Alicea2012,Hung2013,Beenakker2013}. We note that even superconductors with non-zero even BdG Chern number are referred to as topological, but these do not possess Majorana zero modes, and in what follows below, the term topological superconductivity will be limited to those with odd integer $C_{\rm BdG}$.

No naturally occurring topological superconductors of electrons have conclusively been identified yet. One setting in which analogous physics naturally arises is the fractional quantum Hall effect. Many even-denominator fractional quantum Hall states have been observed~\cite{Willett87,Suen92,Shabani09,Shabani13,Kim19,Wang23} and are explained as topological superconductors of composite fermions with $p$-wave, $f$-wave and other kinds of pairings~\cite{Moore91,Read00,Moller08,Faugno20a,Zhao23,Sharma23,Sharma24}.

There have been proposals to engineer topological superconductivity starting from conventional  systems. A notable example is the proximity induced superconductivity at the surface state of a three-dimensional topological insulator by coupling to an ordinary \(s\)-wave superconductor, where the combination of superconductivity with the nontrivial spin-momentum-locked normal-state band structure produces an effective topological superconducting state~\cite{Fu2008,Stanescu2010,Linder2010}. This suggests a general strategy: begin with an electronic band with nontrivial topology and introduce superconducting pairing into it.

LLs provide a particularly natural platform for the realization of this idea because each LL carries a nonzero Chern number. This has motivated extensive work on quantum Hall-superconductor hybrids, in which a quantum Hall system is proximity coupled to a conventional superconductor~\cite{Qi10,Lindner2012,Clarke2013,Mong2014,Zocher2016,Amet2016,Lee17,Jeon2019,Mishmash2019,Chaudhary2020,Kudo2024,Kudo26,Kudo26b}. In such systems, the Abrikosov vortex lattice of the proximity induced superconductor has the same structure as that of the adjacent superconductor. This lattice structure can strongly constrain the allowed topological phases. In particular, for conventional vortex lattices (e.g. hexagonal or square), the degeneracies of the BdG spectrum forces the Chern number to change in even integers, and therefore phases with an odd \(C_{\rm BdG}\) are absent~\cite{Zocher2016,Mishmash2019,Chaudhary20,Schirmer24}.

Another possibility is intrinsic quantum Hall superconductivity, produced when a superconductor is exposed to a magnetic field which generates LLs. Here one considers electrons occupying LLs while interacting with one another with an attractive interaction 
~\cite{tevsanovic1989,chaudhary2021,Schirmer2022a}. 
In this case the superconducting order parameter and the vortex structure are not externally imposed, as is the case for the models mentioned above which consider proximity induced superconductivity, but arise as a part of the solution of the problem.  Schirmer \textit{et al.}~\cite{Schirmer24} studied this problem for spinful electrons with Rashba spin-orbit coupling in a perpendicular magnetic field by obtaining self-consistent solutions of the BdG mean-field equations for an on-site attractive interaction. They found very small regions of topological superconductivity, but showed that introducing a periodic single-particle superlattice potential can produce topological superconductivity for a large range of parameters. The superlattice potential broadens the otherwise narrow Landau-like bands and modifies the vortex structure, with topological superconductivity appearing most strongly when the Abrikosov vortices combine to form ``giant" $h/e$ vortices. This physics would not be possible in a model where superconductivity is proximity induced, and indicates how the physics of a superconductor exposed to a magnetic field can be fundamentally different from that of a proximity induced superconductor.

Disorder is also known to generate topological phases when none are present in the corresponding clean system. Previously it had been shown that disorder can convert a trivial metal into a topological insulator, dubbed topological Anderson insulator~\cite{Li2009,Groth2009, Guo2010, Sttzer2018}. In many studies of proximity induced superconductivity, it has been found that disorder can also reorganize the electronic structure and topology of a superconductor~\cite{Brouwer2011,Sau2013,Borchmann2016,Kudo2024}.  In a two-dimensional spin-orbit-coupled superconductor, disorder can split a direct transition between two clean topological sectors to produce an additional intermediate phase with a distinct Chern number~\cite{Borchmann2016}. More recently, Kudo \textit{et al.}~\cite{Kudo2024} studied disorder in a Rashba-coupled quantum-Hall--superconductor hybrid. In their model the pairing is induced by proximity to a type-II \(s\)-wave superconductor and the BdG Hamiltonian is projected onto a Rashba-coupled LL. Random impurities broaden the LL, and the combined effects of disorder and proximity-induced pairing produce an intervening topological superconducting phase with an odd BdG Chern number. 

In this article, we propose that disorder can also provide a path to topological superconductivity in an intrinsic QH superconductor. We consider spinful electrons with Rashba spin-orbit coupling in a perpendicular magnetic field, interacting through an on-site attractive interaction, and determine the superconducting order parameter self-consistently from the BdG equations in the presence of a random on-site potential. We find that disorder broadens and mixes the otherwise narrow Landau-like bands and can produce odd values of the BdG Chern number. One difficulty in assigning a BdG Chern number is that for a given set of parameters, the BdG Chern number depends on disorder. However, we study an ensemble of disorder configurations and if the same Chern number appears more than 50\% of the time, we assign that Chern number as representing the superconding order that percolates. With that stipulation, we find that disorder produces extended regions of parameter space with odd values of the BdG Chern number, even though the corresponding clean system exhibits even-Chern phases except for certain isolated points. In particular, we find a robust region with $C_{\rm BdG}=-1$ when the chemical potential lies within the disorder-broadened lowest Landau-like band. Our results demonstrate that disorder can generate topological superconductivity in an intrinsic quantum Hall superconductor, where both the superconducting order parameter and its spatial structure arise self-consistently rather than being imposed through proximity coupling.

\section{Model}

We consider spin-$1/2$ fermions on a square lattice in the presence of a perpendicular magnetic field and an on-site potential \(V_i\). The latter is used to model disorder. In our calculations, following standard procedure, the lattice is grouped into a unit cell of size \(N\times M\), containing \(MN\) lattice sites, with the magnetic field chosen such that there is an integer number of flux quanta through this unit cell. We will refer to this as the magnetic unit cell (MUC). For simplicity, in the numerical implementation we take \(V_i\) to be periodic with the same periodicity as the MUC, which allows us to work in momentum space within the reduced Brillouin zone associated with the MUC.

The microscopic Hamiltonian is
\begin{align}
H = H_0 + H_{\rm int},
\end{align}
with
\begin{align}
H_0
&=
\sum_{\langle ij\rangle}
c_i^\dagger e^{iA_{ij}}
\left[
-t\sigma_0
-iV_{\rm so}
(\hat{\mathbf z}\times\hat{\mathbf d}_{ij})
\cdot\boldsymbol{\sigma}
\right]
c_j
+{\rm H.c.}
\nonumber\\
&\quad
+\sum_i (V_i-\mu)c_i^\dagger c_i ,
\label{eq:H0_realspace}
\end{align}
and
\begin{align}
H_{\rm int}
=
-U\sum_i n_{i\uparrow}n_{i\downarrow},
\qquad
U>0.
\label{eq:Hint_realspace}
\end{align}
Here \(c_i=(c_{i\uparrow},c_{i\downarrow})^T\), where \(c_{i\sigma}^\dagger\) (\(c_{i\sigma}\)) creates (annihilates) a fermion of spin \(\sigma\) on site \(i\), \(\sigma_0\) is the identity matrix in spin space, \(\boldsymbol{\sigma}=(\sigma_x,\sigma_y,\sigma_z)\) denotes the Pauli matrices, \(\mu\) is the chemical potential, \(\hat{\mathbf d}_{ij}\) is the unit vector along the directed nearest-neighbor bond from \(i\) to \(j\), and $\hat{\mathbf z}$ is the unit vector normal to the lattice. The parameter \(V_{\rm so}\) denotes the strength of the Rashba spin-orbit coupling. The strength of the on-site attractive interaction is $U$, and $n_{i\sigma}=c_{i\sigma}^\dagger c_{i\sigma}$.

The perpendicular magnetic field is incorporated through Peierls phases in the nearest-neighbor hopping and spin-orbit-coupling matrix elements. The convention is such that the directed sum of \(A_{ij}\) around an elementary plaquette gives the magnetic flux through that plaquette:
\begin{align}
\sum_{\square} A_{ij} = 2\pi \frac{\Phi_p}{\Phi_0},
\qquad
\Phi_0 = h/e ,
\end{align}
where \(\Phi_p\) is the flux per plaquette. The magnetic flux is chosen to be periodic on the \(M\times N\) MUCs defined above. If the total flux through one MUC is \(a\Phi_0\), where \(a\) is an integer, then
\begin{align}
\Phi_p = \frac{a}{MN}\Phi_0.
\end{align}
We neglect Zeeman coupling in order to simplify the calculation, so the perpendicular magnetic field enters only through its orbital coupling.

The disorder potential is generated independently at the \(MN\) sites inside one MUC. We first choose
\begin{align}
\widetilde V_i
\in
\left[
-\frac{W_{\rm dis}}{2},
\frac{W_{\rm dis}}{2}
\right]
\end{align}
from a uniform distribution and subsequently subtract the mean value within the MUC,
\begin{align}
V_i
=
\widetilde V_i
-
\frac{1}{MN}
\sum_{j\in {\rm MUC}}
\widetilde V_j .
\label{eq:disorder_potential}
\end{align}
Thus each disorder realization satisfies
\begin{align}
\sum_{i\in {\rm MUC}}V_i=0.
\end{align}

\subsection{BdG Mean-field decoupling}

Following the usual Bogoliubov--de Gennes mean-field treatment, we decouple the attractive on-site interaction in the pairing channel. We define the on-site spin-singlet superconducting order parameter
\begin{align}
\Delta_i
=
U\expval{c_{i\downarrow}c_{i\uparrow}}.
\label{eq:on-site_order_parameter}
\end{align}
Here, we work at zero temperature, and $\langle\mathcal O\rangle$ denotes the
ground state expectation value of $\mathcal O$.
With this definition, the mean-field interaction is
\begin{align}
H_{\rm int}^{\rm MF}=
-\sum_i
\left[
\Delta_i c_{i\uparrow}^\dagger c_{i\downarrow}^\dagger
+
\Delta_i^\ast c_{i\downarrow}c_{i\uparrow}
\right]
\nonumber\\
&\quad
+
\sum_i\frac{|\Delta_i|^2}{U}.
\label{eq:Hint_mf}
\end{align}
The Hartree contribution from the attractive interaction is neglected in the present calculation. The superconducting order parameter is allowed to vary in both magnitude and phase from site to site within the magnetic unit cell.
The full mean-field Hamiltonian is therefore
\begin{align}
H_{\rm MF}
&=H_0+H_{\rm int}^{\rm MF},
\label{eq:Hmf_realspace}
\end{align}
where $H_0$ is given by Eq.~\eqref{eq:H0_realspace}.

Equation~\eqref{eq:Hmf_realspace} is the starting point for the self-consistent Bogoliubov--de Gennes calculation. In the presence of the magnetic field and disorder, the superconducting order parameter is allowed to adjust independently at every site within the MUC.

\subsection{The Fourier Transform}

Because both the Peierls phase pattern and the on-site potential are periodic on the MUC, it is convenient to label each lattice site by a unit-cell vector \(\mathbf R\) and a sublattice index \(\alpha=1,\dots,MN\) within the MUC:
\begin{align}
i \equiv (\mathbf R,\alpha).
\end{align}
The primitive translation vectors of the MUC are
$\mathbf a_1 = N\,\hat{x}$ and $\mathbf a_2 = M\,\hat{y}$, and we write
$\mathbf R = n_x \mathbf a_1 + n_y \mathbf a_2$
with \(n_x=0,1,\dots,N_x-1\) and \(n_y=0,1,\dots,N_y-1\). Here, \(N_x\) and \(N_y\) are the numbers of magnetic unit cells along the \(x\)- and \(y\)-directions, respectively.

We introduce the Bloch operators
\begin{align}
c_{\mathbf R,\alpha,\sigma}
=
\frac{1}{\sqrt{A}}
\sum_{\mathbf k}
e^{i\mathbf k\cdot\mathbf R}
c_{\mathbf k,\alpha,\sigma},
\label{eq:bloch_transform}
\end{align}
where $A = N_xN_y$ and \(\mathbf k\) belongs to the reduced Brillouin zone associated with the MUC:
\begin{align}
k_x \in \left[-\frac{\pi}{N},\frac{\pi}{N}\right),
\qquad
k_y \in \left[-\frac{\pi}{M},\frac{\pi}{M}\right).
\label{eq:reduced_BZ}
\end{align}
For a finite system with periodic boundary conditions, the allowed discrete momenta may be chosen as
\begin{align}
k_x = \frac{2\pi m_x}{N_xN},
\qquad
k_y = \frac{2\pi m_y}{N_yM},
\label{eq:allowed_k}
\end{align}
with $m_x=0,1,\dots,N_x-1$ and $m_y=0,1,\dots,N_y-1$.
In the thermodynamic limit, the sum over \(\mathbf k\) becomes an integral over the reduced Brillouin zone in Eq.~\eqref{eq:reduced_BZ}.

For each \(\mathbf k\), we first define the spinful single-particle spinor
\begin{align}
\psi_{\mathbf k}
=
\begin{pmatrix}
c_{\mathbf k,1,\uparrow}\\
\vdots\\
c_{\mathbf k,MN,\uparrow}\\
c_{\mathbf k,1,\downarrow}\\
\vdots\\
c_{\mathbf k,MN,\downarrow}
\end{pmatrix},
\label{eq:psik}
\end{align}
and the corresponding Nambu spinor
\begin{align}
\Psi_{\mathbf k}
=
\begin{pmatrix}
\psi_{\mathbf k}\\
(\psi_{-\mathbf k}^{\dagger})^T
\end{pmatrix},
\label{eq:nambu_spinor}
\end{align}
where $(\psi_{-\mathbf k}^{\dagger})^T$ is a transpose of
$\psi_{-\mathbf k}^{\dagger}$.
In this basis, the mean-field Hamiltonian takes the standard form
\begin{align}
H_{\rm MF}
=
\frac{1}{2}
\sum_{\mathbf k}
\Psi_{\mathbf k}^{\dagger}
\mathcal H_{\rm BdG}(\mathbf k)
\Psi_{\mathbf k}
+
E_{\rm const},
\label{eq:Hmf_bdg}
\end{align}
where $\mathcal H_{\rm BdG}(\mathbf k)$ is the \(4MN\times4MN\) BdG matrix, defined by
\begin{align}
\mathcal H_{\rm BdG}(\mathbf k)
=
\begin{pmatrix}
h(\mathbf k)-\mu\,\mathds{1} & \tilde{\Delta}\\
\tilde{\Delta}^\dagger & -[h(-\mathbf k)-\mu\,\mathds{1}]^\ast
\end{pmatrix}.
\label{eq:BdG_matrix}
\end{align}
Here, \(h(\mathbf k)\) is the \(2MN\times2MN\) matrix of $H_0$ in momentum space, excluding the chemical-potential term. The on-site spin-singlet pairing block is given by
\begin{align}
\widetilde{\Delta}
&=
\begin{pmatrix}
0 & -\widehat{\Delta}\\
\widehat{\Delta} & 0
\end{pmatrix},
\label{eq:pairing_matrix}\\
\widehat{\Delta}
&=
{\rm diag}
\left(
\Delta_1^{\rm MUC},
\Delta_2^{\rm MUC},
\dots,
\Delta_{MN}^{\rm MUC}
\right),
\end{align}
where, by the MUC periodicity, \(\Delta_{(\mathbf R,\alpha)}\equiv\Delta_\alpha^{\rm MUC}\) is independent of the MUC position \(\mathbf R\). The antisymmetric spin structure of $\widetilde{\Delta}$ follows from the fermionic antisymmetry of the on-site pair.

The BdG eigenvalue problem is
\begin{align}
\mathcal H_{\rm BdG}(\mathbf k)
\Phi_{n,\mathbf k}
=
E_{n,\mathbf k}
\Phi_{n,\mathbf k},
\label{eq:BdG_eigenproblem}
\end{align}
where we order the eigenvalues as
\(E_{1,\mathbf k}\le E_{2,\mathbf k}\le\cdots\le E_{4MN,\mathbf k}\).
Because of particle-hole symmetry, we have $E_{i,\mathbf k}=-E_{4MN-i+1,-\mathbf k}$.
For a system gapped at zero energy, the first \(2MN\) states therefore
form the negative-energy BdG subspace.

\subsection{Self-consistency equations}

The order parameters must satisfy the mean-field definition in Eq.~\eqref{eq:on-site_order_parameter}. Numerically, this is enforced by diagonalizing \(\mathcal H_{\rm BdG}(\mathbf k)\) on a mesh of \(\mathbf k\)-points in the reduced Brillouin zone, evaluating the on-site anomalous correlation functions from the resulting BdG eigenvectors, and using them to obtain an updated set of order parameters.

The self-consistency equations may be written schematically as
\begin{align}
\Delta^{\rm MUC}_\alpha
=
\frac{U}{A}
\sum_{\mathbf k}
\expval{
c_{-\mathbf k,\alpha,\downarrow}
c_{\mathbf k,\alpha,\uparrow}
},
\label{eq:gap_equation}
\end{align}
To obtain the self-consistent solution, we start from an initial randomly chosen configuration for the order parameter \(\{\Delta^{\rm MUC}_\alpha\}\), diagonalize the BdG Hamiltonian, and calculate the order parameter from Eq.~\eqref{eq:gap_equation}. We then use this as the input configuration, and iterate the procedure until convergence is achieved. Convergence is defined so that the relative difference $|(\Delta_\alpha^{\rm MUC})^{\rm new}
-(\Delta_\alpha^{\rm MUC})^{\rm old}|/|(\Delta_\alpha^{\rm MUC})^{\rm old}|\leq 10^{-5}$ for all $\alpha$. To ensure that the solution is not a local minimum, we ensure that we obtain the same solution starting with many ($\sim 10$) initial configurations. In almost all cases, we find that all initial configurations produce the same final solution. 

For the disordered calculations discussed here, the converged order parameter obtained in the corresponding clean system at the same values of \(U\), \(\mu\), \(V_{\rm so}\), and \(t\) is used as the initial condition. The self-consistency calculation is then repeated in the presence of the disorder potential, allowing the order parameter to adjust spatially within the MUC.

In the phase diagrams below, we characterize the magnitude of superconductivity using
\begin{align}
\Delta_{\max}
=
\max_{\alpha}
|\Delta^{\rm MUC}_\alpha|.
\end{align}
For a converged solution we define the BdG spectral gap as
\begin{align}
E_{\rm gap}
=
\min_{\mathbf k}
E_{2MN+1,\mathbf k}.
\label{eq:spectral_gap}
\end{align}

\subsection{Pairing in the normal-state band basis}

The self-consistent BdG calculation produces the MUC order parameters $\Delta_\alpha^{\rm MUC}$, with $\alpha=1,\dots,MN$, and hence the pairing matrix $\widetilde{\Delta}$ defined in Eq.~\eqref{eq:pairing_matrix}. It is useful to express this pairing in the eigenbasis of the normal-state Hamiltonian $h(\mathbf k)$.

We first diagonalize the normal-state Bloch Hamiltonian,
\begin{align}
h(\mathbf k) \mathbf v_n(\mathbf k)
=
\epsilon_n(\mathbf k)\mathbf v_n(\mathbf k),
\qquad
n=1,\ldots,2MN,
\label{eq:normal_band_eigenproblem}
\end{align}
where the eigenvalues are ordered according to
$\epsilon_1(\mathbf k)\leq\epsilon_2(\mathbf k)\leq\dots\leq\epsilon_{2MN}(\mathbf k)$.
We then collect the normal-state eigenvectors into the unitary matrix
\begin{align}
\mathcal V(\mathbf k)
=
\big(
\mathbf v_1(\mathbf k)\;
\mathbf v_2(\mathbf k)\;
\cdots\;
\mathbf v_{2MN}(\mathbf k)
\big),
\end{align}
so that
the annihilation operators in the band basis are given by
\begin{align}
\gamma_{\mathbf k}
\equiv
\begin{pmatrix}
\gamma_{\mathbf k,1}\\
\gamma_{\mathbf k,2}\\
\vdots\\
\gamma_{\mathbf k,2MN}
\end{pmatrix}
=\mathcal V^\dagger(\mathbf k)
\psi_{\mathbf k}.
\label{eq:gammak}
\end{align}
We then obtain the anomalous part of the BdG Hamiltonian as
\begin{align}
H_\Delta
&=
\frac{1}{2}
\sum_{\mathbf k}
\left[
\psi_{\mathbf k}^\dagger
\widetilde{\Delta}
\left(
\psi_{-\mathbf k}^\dagger
\right)^T
+
{\rm H.c.}
\right]\\
&=
\frac{1}{2}
\sum_{\mathbf k}
\left[
\gamma_{\mathbf k}^\dagger
\Delta^{\rm band}(\mathbf k)
\left(
\gamma_{-\mathbf k}^\dagger
\right)^T
+
{\rm H.c.}
\right],
\end{align}
where $\Delta^{\rm band}(\mathbf k)$ is the band basis representation of the pairing,
whose matrix elements are
\begin{align}
\Delta_{mn}^{\rm band}(\mathbf k)
=
\mathbf v_m^\dagger(\mathbf k)
\widetilde{\Delta}
\mathbf v_n^\ast(-\mathbf k).
\label{eq:projected_pairing_mn}
\end{align}
For notational simplicity, we drop the superscript ``band'' and write this matrix element as $\Delta_{mn}(\mathbf k)$.

Equation~\eqref{eq:projected_pairing_mn} allows us to determine whether the superconducting pairing is predominantly confined to a single normal-state band or whether several bands participate appreciably in the pairing. We focus on the two lowest normal-state magnetic Bloch bands and consider the matrix elements
\(\Delta_{11}(\mathbf k)\), \(\Delta_{12}(\mathbf k)\), and
\(\Delta_{22}(\mathbf k)\). The diagonal elements
\(\Delta_{11}\) and \(\Delta_{22}\) describe intraband pairing within the first and second normal-state bands, respectively, whereas \(\Delta_{12}\) describes interband pairing between them.

The magnitude of each projected matrix element is independent of the arbitrary \(U(1)\) phases chosen for the normal-state Bloch eigenvectors. Indeed, under the transformations
\(\mathbf v_m(\mathbf k)\rightarrow e^{i\chi_m(\mathbf k)}\mathbf v_m(\mathbf k)\), the projected pairing transforms only by an overall phase,
\[
\Delta_{mn}(\mathbf k)
\rightarrow
e^{-i[\chi_m(\mathbf k)+\chi_n(-\mathbf k)]}
\Delta_{mn}(\mathbf k),
\]
and therefore \(|\Delta_{mn}(\mathbf k)|\) is gauge independent. We focus on it in the numerical calculation below.

The antisymmetry \(\widetilde{\Delta}^{T}=-\widetilde{\Delta}\) further implies
\[
\Delta^{\rm band}(\mathbf k)
=
-\left[\Delta^{\rm band}(-\mathbf k)\right]^T,
\]
In particular, the intraband components satisfy
\(\Delta_{nn}(\mathbf k)=-\Delta_{nn}(-\mathbf k)\).
Thus, although the microscopic order parameter is on-site and spin singlet, its intraband projection onto a nondegenerate spin-orbit-coupled band has an effectively odd momentum dependence.

The relative magnitudes of the diagonal and off-diagonal elements of $\Delta^{\rm band}(\mathbf k)$ provide a diagnostic of the validity of a single-band description. If the interband matrix elements are small compared with the intraband pairing, projection onto an isolated band may provide an appropriate reduced description. Conversely, sizeable interband pairing indicates that several normal-state bands must be retained.

We therefore use the band-basis projection as a diagnostic of the momentum structure and multiband character of the self-consistent pairing. The spectral gap and the topological character of the superconducting state are determined from the full $4MN\times4MN$ BdG Hamiltonian.

\subsection{Chern Number}

Once a self-consistent solution is obtained, we compute the BdG Chern number, defined as the total Chern number of all negative-energy BdG bands.

To evaluate the BdG Chern number numerically, we use a standard discretized Brillouin-zone construction. Using the negative-energy eigenstates $\Phi_{n,\mathbf k}$ defined in Eq.~\eqref{eq:BdG_eigenproblem}, we define the overlap $2MN\times 2MN$ matrix as
\begin{align}
\mathcal M_\xi(\mathbf k)_{mn}
=
\Phi_{m,\mathbf k+\delta\mathbf k_\xi}^{\dagger}
\Phi_{n,\mathbf k},
\end{align}
where $\delta\mathbf k_\xi$ is the mesh spacing along $\xi$ direction.
We then define the link variable as
\begin{align}
U_\xi(\mathbf k)
=
\frac{
\det\mathcal M_\xi(\mathbf k)
}{
\left|
\det\mathcal M_\xi(\mathbf k)
\right|
},
\qquad
\xi=x,y.
\label{eq:link_variable}
\end{align}
The field strengh and the BdG Chern number are given by~\cite{fukui2005}
\begin{align}
F_{xy}(\mathbf k)
&=
\arg\!\Big[
\frac{U_x(\mathbf k)
U_y(\mathbf k+\delta\mathbf k_x)}
{U_x(\mathbf k+\delta\mathbf k_y)
U_y (\mathbf k)}
\Big],\\
C_{\rm BdG}
&=
\frac{1}{2\pi}
\sum_{\mathbf k}
F_{xy}(\mathbf k).
\label{eq:chern_number}
\end{align}
An odd integer \(C_{\rm BdG}\) identifies a 
topological superconducting state
with an odd number of chiral Majorana edge modes.

\subsection{Numerical parameters}

The numerical inputs that define one calculation are:
(i) the MUC size \(M\times N\),
(ii) the number \(a\) of flux quanta through the MUC,
(iii) the hopping \(t\), which is set equal to $1$,
(iv) the Rashba spin-orbit coupling \(V_{\rm so}\),
(v) the attraction strength \(U\),
(vi) the chemical potential \(\mu\),
(vii) the disorder strength \(W_{\rm dis}\) and disorder realization,

\begin{figure}[H]
\centering
\captionsetup{justification=raggedright,singlelinecheck=false}
    \includegraphics[width=0.48\textwidth]{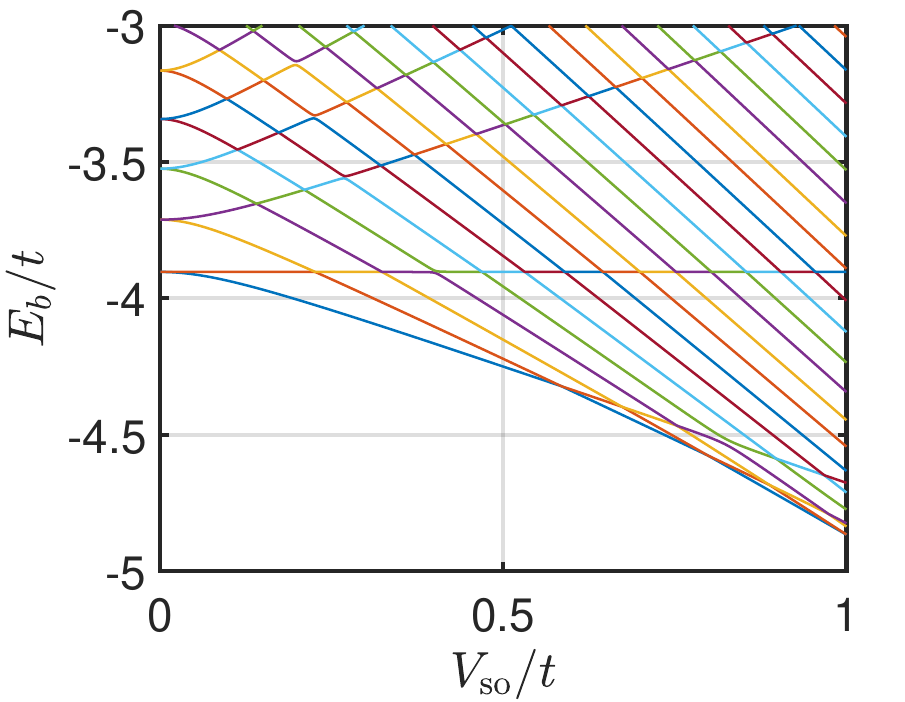}\hfill
    \caption{Normal-state Landau-like bands as a function of Rashba spin-orbit coupling $V_{\rm so}$. The curves show the band midpoint energies $E_{b}$ of $H_0$ for a clean $M=N=8$ magnetic unit cell containing one flux quantum ($\Phi_p/\Phi_0=1/64$). Energies are in units of $t$. We choose to work close to the lowest band at $V_{\rm SO}=0.4t$}
    \label{fig:bands}
\end{figure}

For most of the results given below, we use \(M=N=8\), with one flux quantum through each MUC, so there are \(1/64\) flux quanta per plaquette. Some other values of these parameters have been used to study the robustness of the results. Figure~\ref{fig:bands} shows the evolution of the normal-state Landau-like bands as a function of the Rashba spin-orbit coupling $V_{\rm so}$. Increasing $V_{\rm so}$ splits and reorganizes the Landau-like levels. We choose $V_{\rm so}=0.4t$ for our calculations below, for which the ``lowest Rashba-coupled Landau-like band" can still be clearly identified and followed continuously from weaker spin-orbit coupling. At still larger $V_{\rm so}$, several band crossings occur, making the identification of an individual low-energy Landau-like band less transparent. The value $V_{\rm so}=0.4t$ therefore provides a system with substantial spin-orbit coupling while avoiding these additional band crossings. In the following, we focus on chemical potentials in the vicinity of this lowest band.

\section{Phase Diagram}

\begin{figure*}
\centering
\captionsetup{justification=raggedright,singlelinecheck=false}
    \includegraphics[width=0.9\textwidth]{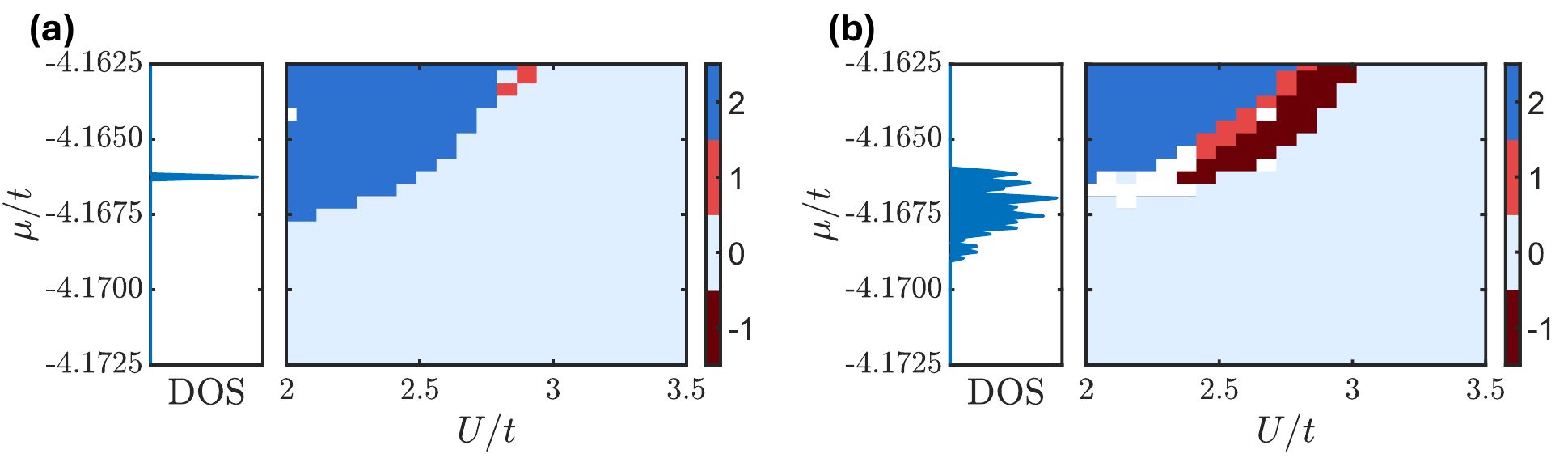}\hfill
    \caption{Phase diagrams of superconductivity (a) without and (b) with disorder as a function of the interaction strength $U/t$ and chemical potential $\mu/t$, where $t$ is the hopping strength. 
    In each panel, the normal-state density of states (DOS) is also shown in a column on the left.
    The calculation is performed for an $8\times 8$ magnetic unit cell containing a total of one flux quantum
    with Rashba spin-orbit coupling $V_{\rm so}=0.4t$. 
    An ensemble of 43 random disorder realizations with $W_{\rm dis}=0.2t$ was studied, and a Chern number (shown in color) was assigned if it occurs for more than 50\% of these realizations. White color displays regions where no Chern number occurs in a majority of the configurations. Evidently, disorder enhances regions in the phase diagram where topological superconductivity with odd BdG Chern number percolates.}
    \label{fig:pdiag}
\end{figure*}

Figure~\ref{fig:pdiag} compares the clean and disordered systems. Figure~\ref{fig:pdiag}(a) shows the normal-state density of states and the BdG Chern-number phase diagram in the clean limit, $W_{\rm dis}=0$. The lowest Landau-like band is very narrow, and the phase diagram is dominated by phases with $C_{\rm BdG}=0$ and $C_{\rm BdG}=2$. A few isolated points with $C_{\rm BdG}=1$ also occur close to the transition between these two regimes.

Figure~\ref{fig:pdiag}(b) shows the corresponding results in the presence of disorder, with $W_{\rm dis}=0.2t$. The random on-site potential broadens the lowest Landau-like band. The superconducting phase diagram is obtained from an ensemble of 43 independent disorder realizations at each available value of $U$ and $\mu$. At each point, we assign a Chern number only when the same value occurs in at least $50\%$ of the disorder realizations; otherwise the point is left white.

As before, there are large regions of superconductors with \(C_{\rm BdG}=0\) and \(C_{\rm BdG}=2\). In the transition region between these two, however, disorder produces sizable regions in which the Chern number is odd: \(C_{\rm BdG}=\pm 1\). These odd-Chern regions 
occur when the chemical potential lies within the energy window associated with the disorder-broadened lowest Landau-like band. The white regions separating some of the colored sectors indicate parameter values for which no single Chern number occurs in a majority of the disorder realizations.

\begin{figure*}
\centering
\captionsetup{justification=raggedright,singlelinecheck=false}
    \includegraphics[width=0.95\textwidth]{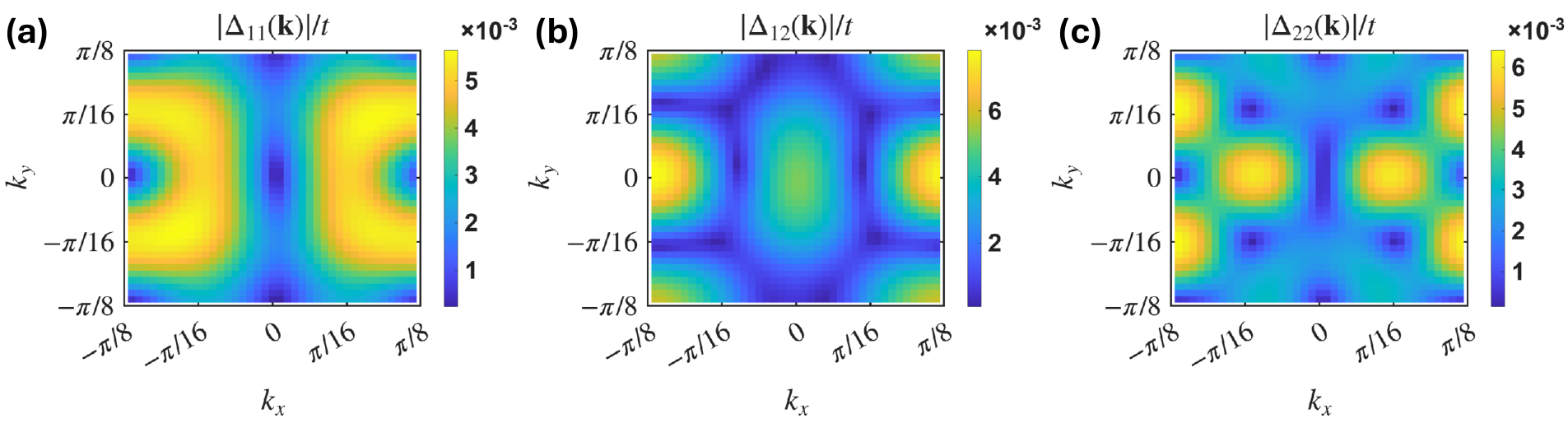}\hfill
    \caption{Magnitude of the pairing matrix elements in the normal-state band basis, defined in Eq.~\eqref{eq:projected_pairing_mn}, for a representative disorder realization in the $C_{\rm BdG}=-1$ phase. (a) $|\Delta_{11}(\mathbf{k})|$, (b) $|\Delta_{12}(\mathbf{k})|$, and (c) $|\Delta_{22}(\mathbf{k})|$. The sizable interband pairing $|\Delta_{12}|$, which is comparable in magnitude to the lowest-band intraband pairing $|\Delta_{11}|$, demonstrates that a theory within the lowest Landau-level band single-band is not adequate for capturing this superconductivity.}
    \label{fig:delta_projected}
\end{figure*}

To gain an understanding of the role of band, or LL mixing, we show in Fig.~\ref{fig:delta_projected} 
\(|\Delta_{11}(\mathbf k)|\),
\(|\Delta_{12}(\mathbf k)|\), and
\(|\Delta_{22}(\mathbf k)|\)
for a representative disorder realization in the
\(C_{\rm BdG}=-1\) phase. The interband matrix element
\(|\Delta_{12}|\) is clearly non-negligible and reaches values comparable to those of the lowest-band intraband pairing
\(|\Delta_{11}|\), indicating that the superconducting pairing in this regime involves multiple low-lying LL bands. A description in which only the lowest LL band is retained is insufficient to describe the physics. 
Also, the zeros or phase structure of \(\Delta_{11}(\mathbf k)\) alone cannot be used to determine the quasiparticle gap closings or the topology of the superconducting state. In particular, a zero of one projected matrix element need not correspond to a zero of the full multiband BdG spectrum.

\section{Discussion and conclusions}
In this paper we have considered the possibility of obtaining topological superconductivity in an intrinsic quantum Hall superconductor through the introduction of disorder. It has long been known that disorder can induce topological insulating phases in systems that are topologically trivial in the clean limit, termed topological Anderson insulator~\cite{Li2009,Groth2009,Guo2010,Sttzer2018}. Our results demonstrate an analogous mechanism in a superconducting system, where disorder can generate regions in the phase diagram with an odd BdG Chern number that are largely absent in the corresponding clean system.

We have studied spinful electrons with Rashba spin-orbit coupling in a perpendicular magnetic field and an on-site attractive interaction, treating the superconducting order parameter self-consistently in the presence of a random on-site potential. The disorder broadens and mixes the otherwise narrow Landau-like bands, and when the chemical potential lies within this disorder-broadened region we find sizeable regions of parameter space with odd values of $C_{\rm BdG}$, including $C_{\rm BdG}=-1$. An ensemble of disorder realizations further reveals regions in which the same odd Chern number occurs for a majority of realizations, suggesting a percolating topological superconducting state. We also find substantial interband pairing in this regime, showing that the odd-Chern phase cannot in general be understood within a projection onto a single Landau-like band. In addition, disorder modifies the self-consistently determined vortex structure, splitting the giant $h/e$ vortex of the clean system into two $h/2e$ vortices. These results provide an example of a topological Anderson superconducting state in an intrinsic quantum Hall superconductor, where both superconductivity and its spatial structure arise self-consistently rather than being imposed by proximity coupling.

{\it Acknowledgments -} 
U.N. was supported in part by the U. S. Department of Energy, Office of Basic Energy Sciences, under Grant no. DE-SC0005042.
K.K. was supported in part by JSPS KAKENHI Grant no. JP25K17318. Both U.N. and K.K. thank the Lodha Theoretical Physics Institute, where part of this work was performed, for hospitality.
We acknowledge Advanced CyberInfrastructure computational resources provided by The Institute for CyberScience at The Pennsylvania State University.

\bibliography{Schirmer.bib,biblio_fqhe.bib}

\end{document}